\documentclass[conference]{IEEEtran}
\IEEEoverridecommandlockouts
\usepackage{newtxmath}

\usepackage[dvipsnames]{xcolor}
\usepackage{tabularx}
\usepackage{multirow}
\usepackage{float}
\usepackage{amsmath}
\usepackage{xcolor,colortbl}
\usepackage[shortlabels]{enumitem}
\usepackage{mathtools}
\usepackage{bm}
\usepackage{tikz-cd}
\usepackage{float}
\usepackage{caption}
\usepackage[labelfont=bf]{caption}
\usepackage{subcaption}
\usepackage{graphicx}
\usepackage{makecell}
\definecolor{salmon}{rgb}{0.945,0.784,0.631}
\definecolor{flarered}{rgb}{0.945,0.251,0.275}
\definecolor{grannysmith}{rgb}{0.659,0.894,0.627}
\usetikzlibrary{matrix,arrows}
\usepackage[version=4]{mhchem} % Formula subscripts using \ce{}
\usepackage{upgreek}
\usepackage{verbatim} % for block comments
\usepackage{physics}
\usepackage{dcolumn}
\usepackage{titlesec}

\usepackage{pgfplots,pgfplotstable}
\usepgfplotslibrary{groupplots}
\usepgfplotslibrary{colorbrewer}
\pgfplotsset{cycle list/Set1}
\usepgfplotslibrary{fillbetween}
\pgfplotsset{compat=1.8}
\pgfplotsset{
    tick align=outside,
    tick pos=left,
    xmajorgrids,
    x grid style={white},
    ymajorgrids,
    y grid style={white},
    axis line style={white},
    axis background/.style={fill=white!89.803921568627459!black},
    legend style={draw=none, fill=none},
    legend cell align=left,
}
\pgfkeys{/pgf/number format/.cd, 1000 sep={\,}}

\pgfplotsset{
    log x ticks with fixed point/.style={
        xticklabel={
            \pgfkeys{/pgf/fpu=true}
            \pgfmathparse{2^\tick}%
            \pgfmathprintnumber[fixed relative, precision=3]{\pgfmathresult}
            \pgfkeys{/pgf/fpu=false}
        }
    },
    log10 x ticks with fixed point/.style={
        xticklabel={
            \pgfkeys{/pgf/fpu=true}
            \pgfmathparse{10^\tick}%
            \pgfmathprintnumber[fixed relative, precision=3]{\pgfmathresult}
            \pgfkeys{/pgf/fpu=false}
        }
    },
    log y ticks with fixed point/.style={
        yticklabel={
            \pgfkeys{/pgf/fpu=true}
            \pgfmathparse{2^\tick}%
            \pgfmathprintnumber[fixed relative, precision=3]{\pgfmathresult}
            \pgfkeys{/pgf/fpu=false}
        }
    }
}
\tikzset{font=\small}

\newcommand*\etal{\textit{et~al }}

\titlespacing{\section}{0pt}{\parskip}{-\parskip}
\titlespacing{\subsection}{0pt}{\parskip}{-\parskip}
\titlespacing{\subsubsection}{0pt}{\parskip}{-\parskip}

\graphicspath{ {./fig/} }

\begin{document}

\title{Micron-scale heterogeneous catalysis with Bayesian force fields from first principles and active learning}

\makeatletter
\newcommand{\linebreakand}{%
  \end{@IEEEauthorhalign}
  \hfill\mbox{}\par
  \mbox{}\hfill\begin{@IEEEauthorhalign}
}
\makeatother

\author{
\IEEEauthorblockN{Anders Johansson}
\IEEEauthorblockA{
\textit{Harvard John A. Paulson School of} \\
\textit{Engineering and Applied Sciences,} \\
Cambridge, MA USA \\
andersjohansson@g.harvard.edu}
\and
\IEEEauthorblockN{Yu Xie}
\IEEEauthorblockA{
\textit{Harvard John A. Paulson School of} \\
\textit{Engineering and Applied Sciences,} \\
Cambridge, MA USA \\
xiey@g.harvard.edu}
\and
\IEEEauthorblockN{Cameron J. Owen}
\IEEEauthorblockA{
\textit{Harvard Department of Chemistry and} \\
\textit{Chemical Biology}, Cambridge, MA USA \\
cowen@g.harvard.edu}
\linebreakand
\IEEEauthorblockN{Jin Soo (David) Lim}
\IEEEauthorblockA{
\textit{Harvard Department of Chemistry and} \\
\textit{Chemical Biology}, Cambridge, MA USA \\
limjs@g.harvard.edu}
\and
\IEEEauthorblockN{Lixin Sun}
\IEEEauthorblockA{
\textit{Harvard John A. Paulson School of} \\
\textit{Engineering and Applied Sciences,} \\
Cambridge, MA USA \\
lixinsun@fas.harvard.edu}
\and
\IEEEauthorblockN{Jonathan Vandermause}
\IEEEauthorblockA{
\textit{Harvard John A. Paulson School of} \\
\textit{Engineering and Applied Sciences,} \\
Cambridge, MA USA \\
jonathan$\_$vandermause@g.harvard.edu}
\linebreakand
\IEEEauthorblockN{Boris Kozinsky}
\IEEEauthorblockA{
\textit{Harvard John A. Paulson School of} \\
\textit{Engineering and Applied Sciences,} \\}
\IEEEauthorblockA{
%\textit{Robert Bosch LLC Research}\\ \textit{and Technology Center}\\
Cambridge, MA USA\\
bkoz@seas.harvard.edu}
}

\maketitle
\thispagestyle{plain}
\pagestyle{plain}

\begin{abstract} % 118/150 words
Quantum-mechanically accurate reactive molecular dynamics (MD) at the scale of billions of atoms has been achieved for the heterogeneous catalytic system of H$_2$/Pt(111) using the FLARE Bayesian force field. This achievement provides accelerated time-to-solution from first principles, with Bayesian active learning enabling efficient and autonomous training of the machine learning model. The resulting model is then deployed in LAMMPS on GPUs using the Kokkos performance portability library. The Bayesian force field provides quantitative uncertainty of predictions on every atomic environment, critical for detecting configurations in large reactive simulations that are outside of the training set. Scaling benchmarks were performed using real-application MD of the \ce{H2}/Pt(111) heterogeneous catalysis on the Summit supercomputer, with simulations reaching 0.5 trillion atoms on 4556 GPU nodes.
\end{abstract}

\section{Justification} %48/50 words max
First micrometer-scale MLFF MD simulations with peak performance of 10.5\ M\,atom-steps/s/node, 70\% improvement over state-of-the-art. MLFF benchmarks of 0.5T atoms, 25$\times$ greater than state-of-the-art. Perfect weak scaling to 40B atoms on 4096 GPU nodes. First uncertainty-aware large-scale reactive MD. Time-to-solution accelerated by orders of magnitude through Bayesian active learning.

%A scalable uncertainty-aware machine learning force field was developed and used in molecular dynamics simulations at $\upmu$-meter length-scales to study a chemically complex heterogeneous reactive system at the scale of billions of atoms. Leading overall time-to-solution \textit{from first principles} is achieved with an automated Bayesian active learning workflow, while establishing state of the art accuracy and scaling for a chemically complex reactive system . Molecular dynamics simulations are implemented on GPUs using the Kokkos library, allowing for massive parallelization on leadership-class compute resources.

\section{Performance Attributes}
%(use a table listing each attribute title and value in a separate row)

%Category of achievement (1+ of:  scalability, time-to-solution, peak performance)

%Type of method used (1 of:  explicit, implicit, both explicit and implicit, semi-implicit, n/a) \textbf{(what is this)}

%Results reported on the basis of (1 of:  whole application including I/O; whole application except I/O; kernel only; other [specify])

%Precision reported (1 of:  single precision, double precision, mixed precision)

%System scale (1 of:  results measured on full-scale system, projected from results of smaller system, other [specify])

%Measurement mechanism (1 of:  timers, FLOP count, static analysis tool, performance modeling, other [specify] )

\begin{table}[H]
\begin{center}
\begin{tabularx}{\columnwidth}{l|l} 
\hline
Performance Attribute & Our Submission \\
\hline
Category of achievement &  Peak performance,  scalability, \\ & time to solution from first principles\\ 
\hline
Performance & 10.5 M\,atoms-steps/s/node \\
\hline
Maximum problem size & 0.5 trillion atoms \\
\hline
Type of method used & FLARE/Kokkos via LAMMPS \\ & molecular dynamics with active learning \\ 
\hline
Results reported on basis of & Whole application including I/O \\ 
\hline
Precision reported & Double precision \\ 
\hline
System scale & Results measured on full-scale system \\ & 4556 GPU nodes (27336 GPUs) \\ & on OLCF Summit \\ 
\hline
Measurement mechanism & Timers in LAMMPS \\ 
\hline
\end{tabularx}
\end{center}
\end{table}

\section{Problem Overview: Elucidating and Predicting the Activity of Heterogeneous Reactions\label{sec:problem}}
%1 p max, description of the problem and its importance, in terms understandable to a non-specialist

Interface reactions play a crucial role in determining the performance and durability of functional materials, including corrosion of metals, tribology of lubricated surfaces, mechanical properties of alloys and composites, transport across electrolyte-electrode interfaces in batteries, and heterogeneous catalysis. Heterogeneous catalysis lies at the core of sustainable chemical production, and rational catalyst design is required to move beyond the conventional trial-and-error approaches \cite{Personick16p20150077}. In this regard, elucidating surface reactivity is central to improve catalytic performance. But surfaces can undergo restructuring under reactive environments. Such dynamical processes necessitate elucidation and atomistic modeling of the underlying mechanisms and timescales.

Interfacial phenomena involve large length-scales and long timescales. Computational modeling at such scales presents a massive challenge, but also an incredible opportunity for combining state-of-the-art (SOTA) methods development with leadership-class computing resources. \textit{Ab initio}, or first principles, methods based on the explicit quantum mechanical treatment of the electronic structure, such as density functional theory (DFT), are essential for establishing the structure-activity relationship of catalytic surfaces \cite{Hammer2000TheoreticalSS}. However, their poor scaling and high computational cost ($O(N_\mathrm{e}^3)$ with the number of electrons) preclude their use for large-scale and long-timescale simulations. First-principles modeling of catalytic systems are often limited to static configurations or fast high-temperature reactions on small periodic structures \cite{doi:10.1073/pnas.1612106114} or nanoclusters \cite{doi:10.1021/acsnano.6b07409}. Such simulations typically span picoseconds and nanometers, far behind the microsecond and micrometer time- and length-scales of real catalysts.

For complex surface structures with a wide variety of dynamic reactive sites, it is highly nontrivial to replace \textit{ab initio} methods with cheaper and more scalable methods. Previously, cluster expansions \cite{C7CP02221B} and empirical \cite{Senftle2016} / machine-learned (ML) \cite{Gerrits_2021} force fields have been employed. These methods rely only on the atomic positions in local environments and do not explicitly consider the electronic structure, thereby scaling linearly with the number of atoms $O(N_\mathrm{atom})$ and providing several orders-of-magnitude acceleration compared to DFT.

However, the accuracy, transferability, and speed of these models have remained limited for applications in computational catalysis. One such example can be found in the modeling of \ce{H2} reactivity on transition metal surfaces, a prototypical system in heterogeneous catalysis. \ce{H2} reactivity has been a subject of extensive experimental and computational investigation due to its industrial importance in numerous catalytic processes (e.g., selective hydrogenation \cite{Luneau20p12834} and hydrogen storage \cite{Pyle18p19098}). On noble metals such as Pt, an \ce{H2} molecule approaches the Pt surface and dissociates to form two chemisorbed H atoms, which diffuse and eventually recombine and desorb from the surface as molecular \ce{H2} (Figure~\ref{Fig1}) \cite{montano2006hydrogen}.

\begin{figure}[h]
    \centering
    \includegraphics[width=0.48\textwidth]{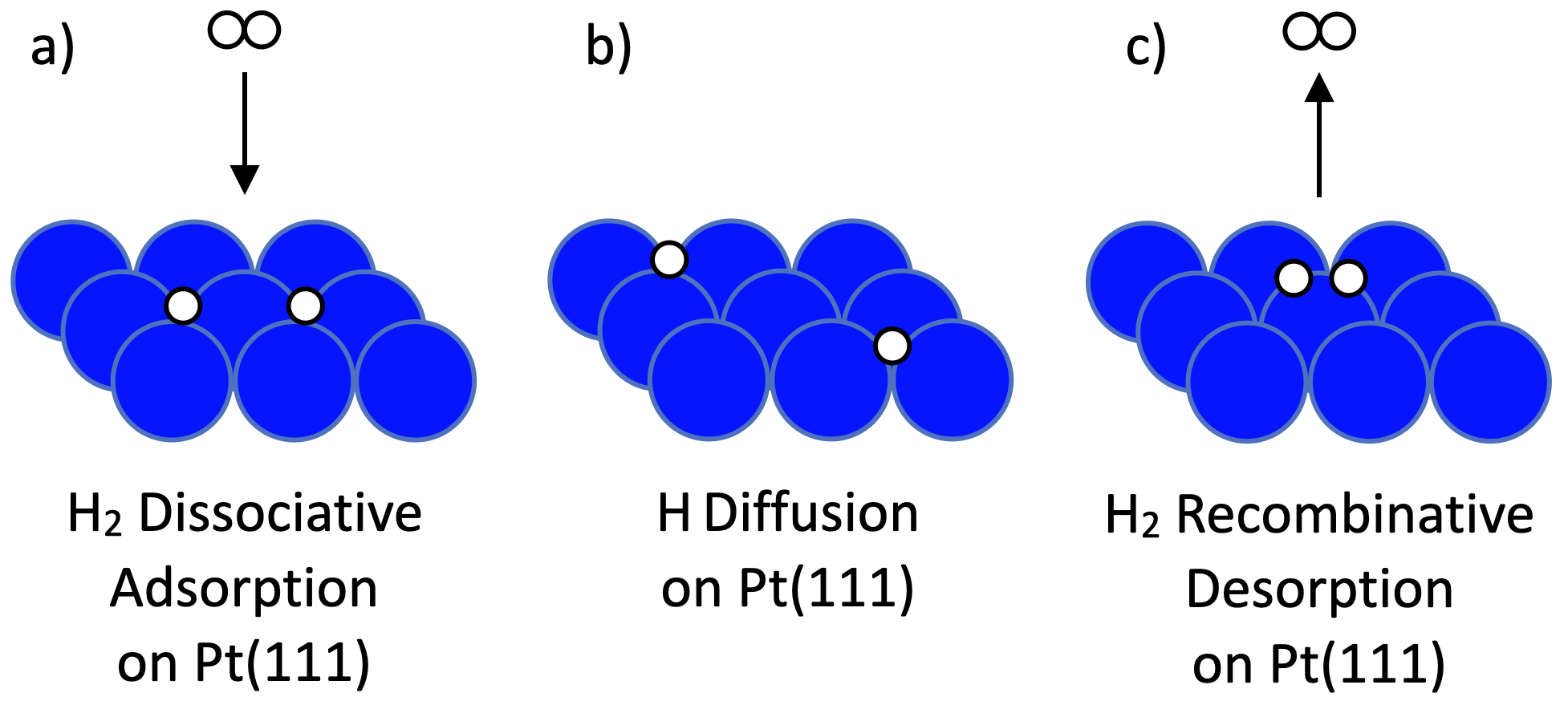}
    \caption{Mechanisms of \ce{H2} reactivity on Pt(111) surface. (a) Dissociative adsorption (chemisorption) of \ce{H2} on Pt(111); (b) diffusion of atomic H on Pt(111); (c) recombinative desorption of \ce{H2} from the surface.}
    \label{Fig1}
\end{figure}

A variety of force field-based methods have been employed to model H/Pt, most notably ReaxFF \cite{Gai16p9780}, embedded-atom method (EAM) \cite{Tokumasu12p1144}, tight-binding \cite{Ahmed11p10503}, low-dimensional models \cite{Kroes16p3658}, and ML force fields \cite{Gerrits_2021}. Empirical models such as ReaxFF and EAM have been shown to be inaccurate and unreliable for high-temperature long-timescale simulations \cite{vandermause2021active}. Several challenges impede the applicability of available models. The first challenge for reactive force fields is the computational cost that increases with model complexity, which limits the size and time scales of simulations. Second is the limited expressiveness of their parametric functional forms, which lower their accuracy and transferability: they only consider either a bare surface interacting with a single \ce{H2} molecule or H-covered surface coupled to an implicit reservoir representing the gas phase, failing to model ambient pressure conditions relevant to real-life catalysis, such as the Haber-Bosch process which operates at pressures well above 100~bar. Such conditions involve multiple gas-phase \ce{H2} molecules and chemisorbed H atoms on the surface. Third, none of the existing reactive models provide any principled metric of uncertainty or reliability. Uncertainty quantification is critical for evaluating model reliability for configurations that may be outside of the training set. This is especially critical for simulations involving rare transition or migration events that occur in chemical reactions. Finally, force field models typically require laborious \textit{ad hoc} manual enumeration of structural configurations for generating the training set.

The challenges outlined above hold generally for the modeling of reactive systems, not just the particular H/Pt system. Hence, progress in computations of complex reactive systems necessitates the development of flexible models for force fields that are {(1) scalable and fast, reaching large sizes and long timescales of simulations needed to capture reaction kinetics, (2) chemically accurate and transferable to a range of reaction conditions and complex chemical environments, (3) endowed with predictive uncertainty quantification and reliability guarantees, and (4) quickly and autonomously trained}.

\section{Current State of the Art\label{sec:sota}}
%vasp active learning (non-mapped GP and small system sizes?), DeepMD (DPgen), SNAP, MTP, GAP 

%\textit{quantitative discussion of current SoA, with emphasis on performance-related aspects}

% three parts:
% AIMD trajectory: see the article. Water splitting, small nano-particle. (% reference 315-319, 351, 352)
% Cluster expansion and classical force field: general catalysis system. then another paragraph on narrowing down the specific H/Pt system
% The section 5.4.2 of the published paper
% There have been attempts with machine learning force fields for catalysis system. Some of them do pure alloy in a vacuum (log of citations, from the table), or CuAu + water with DeePMD but slow
% Training data size: emphasize that SNAP and DeePMD use a lot of data

% FLARE is designed for the challenge of heterogeneous catalysis. We need to discriminate this work from previous machine learning potentials like SNAP and DeePMD to avoid too much similarity
% We have bulk that can be handled well by classical force field, but here we have surface and molecules, with different bond natures and vibrational modes, different parts of the system has different characteristics. The molecules have high frequency and requires much smaller time step, while the bulk has much slower dynamics. Therefore, needs much longer time to integrate. We can not use the methods for pure bulk for this heterogeneous system. 
% FLARE make this catalysis application improved by xxx in accuracy (compared with classical ff) and speed (compared with AIMD)

Over the last decade, ML force fields have emerged as a promising path to highly accurate and scalable atomistic simulations. 
Notable developments include Behler-Parrinello neural-network potentials (HDNNP) \cite{behler2007generalized}, Gaussian approximation potentials (GAP) \cite{bartok2010gaussian}, spectral neighbor analysis potentials (SNAP) \cite{thompson2015spectral}, moment tensor potentials (MTP) \cite{shapeev2016moment}, ANI neural-network potentials \cite{smith2017ani}, SchNet \cite{schutt2017schnet}, DeePMD \cite{zhang2018deep}, atomic cluster expansion (ACE) \cite{drautz_atomic_2019}, and NequIP \cite{batzner2021se}.
These MLFFs can in principle make predictions at near \textit{ab-initio} accuracy, but none of the previous models simultaneously address all four criteria listed in Section III. Here we discuss the current state of the art in the four aspects separately.

\subsection{Scalability and Speed \label{subsec:sota_performance}}

Two existing descriptor-based MLFF models demonstrated the capability to simulate billions of atoms, but only describing dynamics of periodic bulk systems. Importantly, none of the existing models describe reactive dynamics, in either heterogeneous or even homogeneous systems.
The SNAP model was employed on bulk carbon system with peak performance 6.21 M atoms$\cdot$steps/s/node \cite{nguyen2021billion}.
Likewise, DeePMD achieved 0.3 M atoms$\cdot$steps/s/node for homogeneous bulk systems - water and copper - on the OLCF Summit machine \cite{jia2020pushing}, with the performance recently improved to 2.0 M atoms$\cdot$steps/s/node \cite{guo2022extending}. Other models, like graph neural networks, have not been shown to approach such scales due to difficulty in parallelization. 
On Summit, DeePMD and SNAP are able to simulate 3.9 and 20 billion atoms, respectively, using 4560 and 4650 GPU nodes, as listed in Table \ref{tab:radar}.%% DeePMD  was also benchmarked on 9936 CPU nodes on the Fugaku supercomputer, reaching 25 billion atoms.

\subsection{Accuracy and Transferability}

As mentioned in section III, existing reactive force fields based on either empirical or ML models suffer from transferability and accuracy issues. Specifically for the H/Pt heterogeneous systems, previous force fields are limited to describing low-pressure and low-temperature conditions. Latest models include a ReaxFF model from Gai \etal~\cite{Gai16p9780} and a HDNNP from Gerrits \cite{Gerrits_2021}. Both models were optimized to match DFT calculations of H-covered Pt surfaces. The ReaxFF model was designed to captures H adsorption behavior, while the HDNNP can model a single \ce{H2} molecule from a molecular beam interacting with the surface. However, from our benchmarks~\cite{vandermause2021active}, models constructed with only low-energy structures without an explicit gas-surface interface do not have sufficient accuracy in describing the ambient pressure environment, and thus cannot thermally activated states of the full catalytic reaction pathway.

\subsection{Uncertainty Quantification at Scale\label{subsec:uncertainty}}

Bayesian models such as Gaussian processes (GPs) provide both mean and variance predictions, where the variance can be used to quantify uncertainty.
But the computational cost of predicting the variances scales with the training set size as $O(n_\mathrm{S}^2)$, more expensive than the $O(n_\mathrm{S})$ scaling of predicting the mean.
This inferior scaling applies to the Bayesian regression approach used by Jinnouchi \etal \cite{jinnouchi2019phase}, limiting the application to short time scales and small length scales. Podryabinkin \etal used D-optimality as an indicator of uncertainty of a linear descriptor-based model~\cite{podryabinkin2017active}, which provides only a qualitative decision boundary but not quantitative uncertainty for each structure. 
In the class of neural network models, such as DeePMD, the ``ensemble'' technique is used to approximate uncertainty by the variance of predictions coming from multiple light-weight models \cite{zhang2020dp}. This approach requires additional computational resources for multiple neural networks, and a dataset of several thousand structures, while yielding a final model still without uncertainty.
In an earlier version of the FLARE model with 2+3-body kernels \cite{vandermause2020fly,xie2021bayesian}, we verified uncertainty accuracy and circumvented this scaling problem by an exact mapping of the GP mean and variance to spline functions, reducing the prediction cost of energies and forces, together with their uncertainty, to a cost comparable with empirical 3-body force fields. This represents the SOTA, which we improve on using many-body kernels and sparse GPs\cite{vandermause2021active}, as described in Section \ref{sec:innovation}.

\subsection{Time-to-solution from first principles}
Time-to-solution for MLFF has previously been defined as the production simulation speed of the finally trained model, i.e. the M\,atom-steps/s/node \cite{jia2020pushing,nguyen2021billion}.
However, the cost of generating relevant and sufficiently diverse first-principles reference data and training the force fields can take months and requires domain knowledge and arduous human trial and error.
For example, without uncertainty quantification and active learning, the SNAP potential of carbon is trained from static DFT and ab-initio MD data \cite{willman2020quantum}, which takes days or weeks to generate. DeePMD used a relatively large dataset of 7646 frames of DFT data for a homogeneous pure bulk copper model \cite{zhang2020dp}. Training of empirical force fields, like ReaxFF, relies on manual selection of relevant atomic configurations and typically takes months of human iteration. The generation cost increases rapidly with structural and chemical complexity, as well as rare events like chemical reactions. Therefore, ``time-to-solution \textit{from first principles}'' in this work emphasizes the entire data workflow starting from system description, selecting training configurations, performing reference DFT calculations, and training the final model. 

\textit{Active learning} workflows can replace this arduous human-driven process. In this approach, molecular dynamics or another sampling method is used to explore the configuration space with the fast surrogate model, and expensive \textit{ab initio} methods are only called when the model uncertainty exceeds a threshold. To actively train their corresponding models mentioned above Zhang \etal used the ensemble method to train DeePMD force fields~\cite{zhang2020dp}, Podryabinkin \etal used D-optimality~\cite{podryabinkin2017active}, and Jinnouchi \etal used Bayesian variance~\cite{jinnouchi2019phase}. As these models do not produce large-scale MD with uncertainties, no further iterative model refinement is possible. Our earlier FLARE 2+3-body model was used to perform hierarchical active learning where mapped quantitative uncertainty of sequentially larger-scale MD was used to detect configurations outside the training and to refine the model \cite{xie2021bayesian}. In this work, we improve upon this SOTA approach by implementing many-body descriptor based kernels \cite{xie2022uncertainty} as described in Section \ref{sec:innovation}.

\section{Innovations Realized \label{sec:innovation}}
%\textit{what the innovations are and how they were achieved}

% In this work, we have developed a multi-tier approach to improve the scalability and accuracy of reactive FFs. In the first tier, we employ Bayesian active-learning to \textbf{decrease time-to-solution} (defined in Section IV D), where machine-learned force fields using Gaussian Process(GP) are trained on-the-fly during molecular dynamic simulations from \textit{ab initio} methods. The expensive ab initial methods are only sporadically called when encountering atomic environments with high uncertainty.\cite{vandermause2021active}
% In the second tier, we accelerate the resulting GP onto a much faster parametric? model (mapped-GP), with O(N) scaling.
% In the third tier, we take advantage of current mass-introduction of GPU resources by making this mapped-GP model GPU compatible, and further improve scaling and speed by using the Kokkos portability library for performing molecular dynamics within LAMMPS.

In this work, we employ the FLARE framework to develop a fast and accurate Bayesian many-body force field and simulate the heterogeneous H/Pt catalytic reaction, demonstrating significant improvements over the SOTA performance in the following aspects:
\begin{enumerate}
    \item The speed of our model exceeds both the non-reactive DeePMD benchmark on simple thermal dynamics in homogeneous systems by $30\times$ for bulk water \cite{jia2020pushing} and $4\times$ for bulk copper \cite{guo2022extending}, the SNAP benchmark by $1.7\times$ for bulk carbon \cite{nguyen2021billion}, and the reactive force field ReaxFF by at least $5\times$. ReaxFF benchmark was on 1M atoms on one V100 GPU node, and does not scale linearly with the system size. (Section\ref{subsec:bff} and \ref{subsec:kokkos})
    
    \item The accuracy of FLARE exceeds the only available model for H/Pt heterogeneous catalysis reactions, ReaxFF, by an order of magnitude in the mean absolute error of force predictions relative to first-principle calculations, also evidenced by nonphysical structural evolution observed with ReaxFF at realistic chemical conditions \cite{vandermause2021active}.
    
    \item Predictive uncertainty quantification is implemented in FLARE with minimal impact on speed, even in 1B-atom scale MD calculations. This capability is unique among any available machine learning potentials. (Section\ref{subsec:bff})
    
    \item The time-to-solution from first principles is 214.8 wall time hours, and 568.8 node hours realized by fully autonomous Bayesian active learning, which outperforms other machine learning potentials trained with AIMD data \cite{willman2020quantum} by orders of magnitude. The number of first-principle calculations used for our training is $13\times$ smaller than the DeePMD copper potential from DP-GEN \cite{zhang2020dp}, demonstrating a higher data efficiency. (Section\ref{subsec:bal})
\end{enumerate}

With the above implementation, we are able to model the heterogeneous H/Pt catalytic process at micrometer length scales under realistic reaction conditions, making this the largest reactive MD simulation to date.

\subsection{Bayesian Active Learning\label{subsec:bal}}

Utilizing the inherent uncertainty quantification of the FLARE Bayesian force field (BFF), we design the Bayesian active learning approach, enabling orders of magnitude higher efficiency in data generation than using ab-initio MD \cite{vandermause2020fly,vandermause2021active}, as detailed in Table \ref{tab:training_stats}.
As illustrated in Figure~\ref{fig:otf}a, the MD simulation is driven by the fast Bayesian force field, with the uncertainty calculated at each step. When the uncertainties are greater than a decision threshold $\Delta_\text{DFT}$, indicating that the current structure is significantly different from those stored in the training set, the MD is interrupted and DFT is called on the uncertain structure. The results of that DFT calculation are then added to the training set, and the Bayesian force field model is updated.

At the beginning of the training procedure, when the training set is small, DFT needs to be called frequently. As the training set grows, DFT is called less frequently, until the simulation has sufficiently explored the configuration space, meaning that a diverse training set is automatically created.
Due to the high data efficiency of the GP, we can obtain a trustworthy model for MD simulations, while the uncertainty quantification helps to detect unexplored region and improve the model accuracy.

\begin{figure}[htbp]
    \centering
    \includegraphics[width=0.45\textwidth]{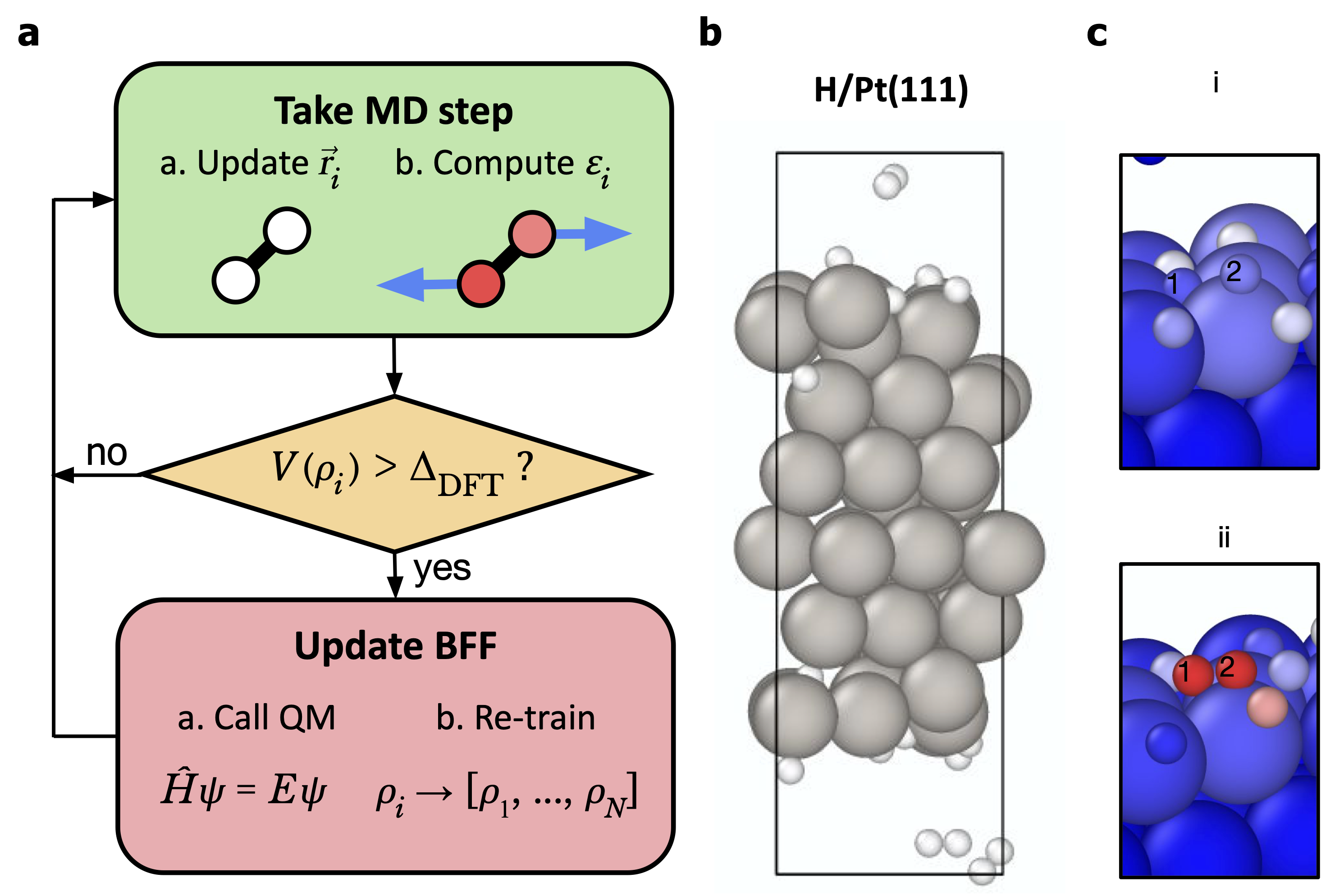}
    \caption{a. Bayesian active learning workflow driven by predictive uncertainty. b. H/Pt(111) configuration for training. c. Atoms colored by uncertainty during the training. \cite{vandermause2021active}}
    \label{fig:otf}
\end{figure}

\subsection{Acceleration of the Bayesian Force Field\label{subsec:bff}}
%\subsection{FLARE Bayesian Force Fields from Sparse Gaussian Process}

In this work, we use the FLARE Bayesian force field framework, a descriptor-based model with not only the energy, forces and stress, but also the uncertainty quantification associated with the predictions \cite{vandermause2021active}. 
%\ptitle{What is atomic environment. What is GP. what is kernel}
The model is constructed with three parts: the descriptors, the kernel, and the Gaussian process regression model. 

\textit{Local structure descriptors.} A local atomic environment $\rho_i$ of an atom $i$ consists of all the neighbor atoms within a cutoff radius $r_\text{cut}$ and includes information of atomic positions and chemical species. 
Atomic cluster expansion (ACE) B2 descriptors \cite{drautz_atomic_2019} are used to describe the local atomic environments, constructed from an equivariant basis of spherical harmonics and radial functions. We denote the normalized ACE-B2 descriptors as $\boldsymbol{d}_i=\mathbf{d}_i/\|\mathbf{d}_i\|$.

\textit{Kernel.} The kernel function $k(\rho_i, \rho_j)$ quantifies the similarity between two atomic environments. We use the normalized inner product kernel: $k_\xi(\rho_1, \rho_2) = \sigma^2 \left(\boldsymbol{d}_1 \cdot \boldsymbol{d}_2 \right)^\xi$, where $\sigma^2$ is the signal variance optimized by maximizing the log likelihood of SGP, and $\xi$ is the power of the inner product. 

\begin{figure}[htbp]
    \centering
    \includegraphics[width=0.48\textwidth]{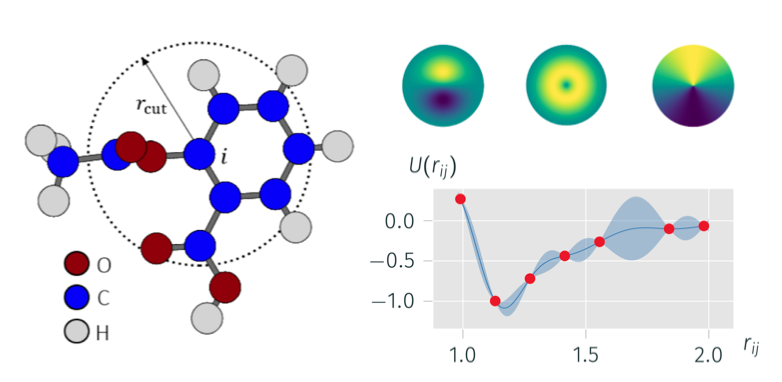}
    \caption{Left: local atomic environment. Upper right: examples of spherical harmonics used for the descriptors. Lower right: GP regression (schematic).}
    \label{fig:env_gp}
\end{figure}

\textit{Mapped Sparse GP.} We use Sparse Gaussian process (SGP) regression, assuming a Gaussian joint distribution of all the training data and test data, to predict the energy/forces/stress and the corresponding uncertainties for each atomic structure as the mean and variance of the posterior distribution,
\begin{align}
    \varepsilon(\rho_i) &= \sum_{s\in S} k_2(\rho_i, \rho_s) \alpha_s = \sum_{s\in S} \boldsymbol{d}_i^\top \boldsymbol{\beta} \boldsymbol{d}_i\,, 
    \label{eq:sgp_mean}\\
    V(\rho_i) &= \sigma^2 k_1(\rho_i, \rho_i)
- \sigma^4 \sum_{s,t\in S} k_1(\rho_i, \rho_s) (K_{SS}^{-1})_{st} k_1(\rho_t, \rho_i)\ \nonumber \\
&= \boldsymbol{d}_i^\top \boldsymbol{\gamma} \boldsymbol{d}_i\,,
\label{eq:sgp_var}
\end{align}
where the $\alpha_s$ are constants computed from training labels and kernels between training data, $S$ is the set of sparse representatives of the training data set, and $K_{SS}$ is the kernel matrix between the environments in $S$. The uncertainties are given by the square root of the variance. As a trade-off between accuracy and performance, we use the $\xi=2$ kernel for mean and the $\xi=1$ kernel for variance evaluations (see detailed discussions in \cite{vandermause2021active,xie2022uncertainty}). 

As shown in Eq.\eqref{eq:sgp_mean}, the cost of energy/forces/stress predictions scales as $O(n_S n_d)$, and the uncertainty predictions as $O(n_S^2 n_d)$, where $n_d$ is the descriptor dimension, and $n_S$ the sparse training set size. 
For complex systems, a large number of training data are usually needed for sufficient accuracy, indicating $n_S \gg n_d$, which makes the naive evaluation of the SGP model slow.
In FLARE, we introduced a key unique way of reducing the scaling of the mean and variance to $O(n_d^2)$ \cite{vandermause2021active,xie2022uncertainty}. 
Specifically, with inner product kernels, a highly efficient but lossless exact mapping is available via reorganization of the summation, as shown in Eq.~\eqref{eq:sgp_mean}.
The $\boldsymbol{\beta}$ and $\boldsymbol{\gamma}$ tensors can be computed from the training data descriptors once and stored, and during the prediction we directly evaluate the quadratic model without the explicit summation of all sparse data $S$, making the computational cost independent of the training data set size $n_S$.

\textit{Acceleration of force predictions.} In SGP, the evaluation of forces requires computing the gradient of the descriptors $\nabla \mathbf{d}_i$ and is the bottleneck of the performance. 
We observe that in the mapped quadratic model, this can be circumvented by reordering the summation, such that the cost of the forces is reduced by a factor of $n_\text{max}$, the number of radial basis functions in the descriptors. This provides an overall speedup of approximately $2\times$ compared to calculating forces directly from $\nabla \mathbf{d}_i$. 

\textit{Acceleration of heterogeneous multi-element system.} The dimension of the ACE-B2 descriptors grows as the square of the number of chemical elements.
If an atom only has neighbors of its own chemical species, such as in the Pt bulk or in the \ce{H2} gas, the majority of the descriptor components will be zero. By checking the neighborhood of each atom before the descriptor calculation, we avoid computing unnecessary components and drastically reduce the size of the quadratic model in Eq.\eqref{eq:sgp_mean}. For all atoms not near the surface, the dimension of the descriptor is reduced from 420 to 112, thus accelerating the quadratic model by more than an order of magnitude.

\subsection{Hardware acceleration with Kokkos\label{subsec:kokkos}}
Over the past few years, hardware accelerators such as graphical processing units (GPUs) have seen widespread adaptation in high-performance computing (HPC) centers around the world, with 7 out of 10 most powerful supercomputers in the world being GPU-based \cite{top500}. GPUs offer massive parallelism with a large number of cores, making them ideal for MD simulations where the forces and energies of large numbers of atoms need to be computed.

The Kokkos performance portability library~\cite{kokkos} is developed for hardware-agnostic programming, which automatically fits different architectures and programming models such as OpenMP, CUDA and HIP during compilation. 
Kokkos has been tightly integrated with the Large-scale Atomic/Molecular Massively Parallel Simulator (LAMMPS) \cite{thompson2022lammps}, such that MD runs entirely on GPUs with zero CPU-GPU data transfer except for I/O processes. 
All CPU-GPU and GPU-GPU communication is handled by LAMMPS's well-proven routines, giving us easy access to excellent performance and scalability.

We integrate the FLARE force field with LAMMPS and optimize the performance by making extensive use of Kokkos's hierarchical parallelism functionality, which is directly mapped to CUDA blocks, warps and threads. 
For most computational steps, one CUDA block of threads will work together on one atom, while the intra-block parallelism is exploited for an atom's neighbors and/or the components of its descriptor vector.
For example, each CUDA-block performs one \(\boldsymbol{\beta} \boldsymbol{d}_i\) matrix-vector product in Eq.~\eqref{eq:sgp_mean}, and each warp of threads computes one component of the result with a parallel dot-product between one matrix row and the descriptor vector. The repeatedly accessed descriptor vector is loaded into the highest-performance shared memory on the GPU. 
The matrix-vector product is computed in chunks whenever the full-sized descriptor vector is too large to fit into shared memory.

The multi-level parallelism enables saturated performance on GPUs for small systems of hundreds of atoms and extreme-scale systems with billions of atoms, thus utilizing the full power of GPUs for both long time-scale and large length-scale simulations. 
% For small systems, the performance is usually unsaturated with empirical potentials.
% This multi-level parallelism has the benefit of greatly extending the viability of GPUs for MD by ensuring that the GPU will be saturated for fairly small systems. Many applications of MD simply do not require millions or billions of atoms, thus they would not saturate the GPU and its massive parallelism. With our hierarchical approach, system sizes down to less than one thousand atoms will still get the full benefit of GPUs, thus enabling longer timescale simulations.
% On the other end of the spectrum, FLARE can accommodate extreme-scale simulations. In principle, 
Descriptor-based methods like FLARE are memory-hungry because the derivative of every descriptor component of every atom with respect to each neighbor is required. 
The issue is avoided by dividing atoms into batches, so that only a subset of descriptors and gradients are stored at a time. The batch size is automatically calculated based on memory. 
Due to the multi-level parallelism, peak performance is achieved even for small batch sizes, thus LAMMPS's neighbor lists ultimately become the limiting factor in terms of system size.
%The performance of FLARE force field implemented with Kokkos is $60\times$ (?) on 1 GPU faster than the implementation without Kokkos on 1 CPU.

\section{How Performance Was Measured\label{sec:perf}}
%(Note that preference is given to performance actually measured [not projected], based on the entire application [including I/O] and with uniform precision.  Explain in detail if any portion of total runtime was not included in the measurements, if and where different precisions were used, or any attributes listed in Section 3 as “other”).

\subsection{Scientific applications used to measure performance}

As discussed in Section \ref{sec:problem}, it is critical to model catalytic reactions at conditions relevant to real-life catalysis.
To model the prototypical \ce{H2} reaction at ambient pressures,
we consider a combined system of H$_2$ gas interacting with a Pt(111) slab (Figure \ref{fig:ptH}). The Pt surface has 885 million atoms with the area of 2.2 $\upmu\mathrm{m} \times$ 1.9 $\upmu\mathrm{m}$, while 120 million H atoms are scattered in the space, yielding a total of 1 billion atoms.

Such large system sizes allow for meaningful direct two-phase reaction simulations at experimentally relevant conditions.
Our previous simulations show that a smaller supercell with 448 H atoms only observes twenty total reactions (splitting and recombination) within 500 ps at 450 K.\cite{vandermause2021active}
The limited number of events is not sufficient for statistical analysis of reactivity. This problem can be mitigated by drastically increasing the length-scale with a larger Pt surface and 120 million H atoms, allowing more reactions to happen in a short period of time and providing sufficient statistics for studying the reaction rates and mechanisms.

\begin{figure}[htbp]
    \centering
    \includegraphics[width=0.45\textwidth]{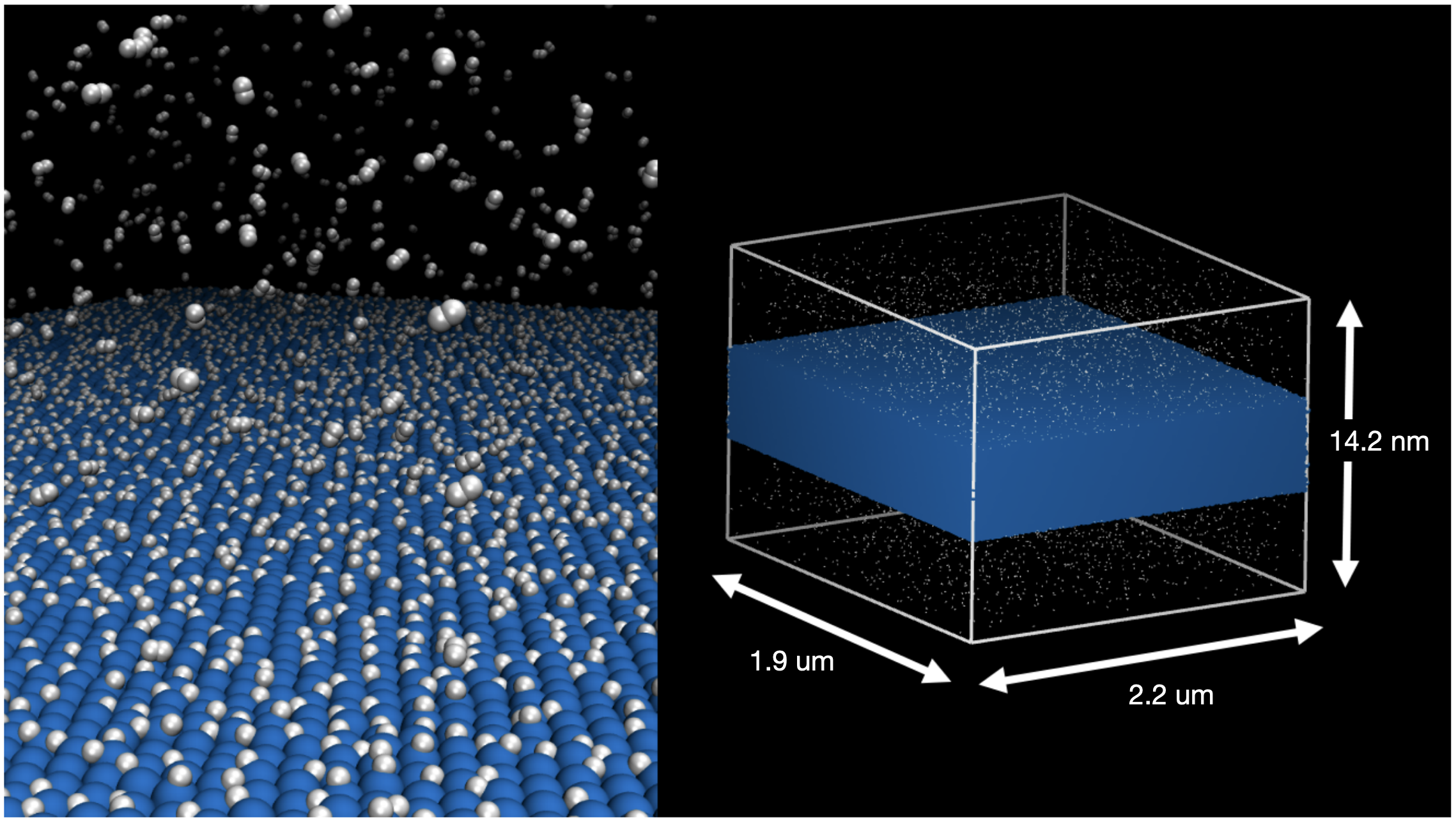}
    \caption{Perspective view of the 1B-atom system and its dimensions. The height of the simulation box is not shown to scale.}
    \label{fig:ptH}
\end{figure}

The large-scale H/Pt simulations were prepared in three steps. A smaller supercell with 216000 Pt and 29200 H atoms is first generated by randomly assigning \ce{H2} molecules to the vacuum and a 0.8 monolayer of \ce{H} atoms adsorbed on the Pt surface.
This supercell was thermalized at 450 K for 2.1 ns till the surface coverage of \ce{H} approached equilibrium.
The supercell was then replicated to yield the final structure. 

Catalytic reactions were modeled at 450 K using the Nos\'e-Hoover thermostat. A time step of 0.1~fs was employed to account for the much faster vibrational mode of H compared to Pt. Snapshots of the trajectory were dumped every 0.1 ps, allowing for post-processing of the data to determine the number of reactions observed. Hydrogen coordination numbers were determined to be 1 or 0 whether another hydrogen atom was within 1 \AA, where a change in this coordination number would correspond to a reaction event taking place (e.g., 1 $\to$ 0 being dissociation \& 0 $\to$ 1 being recombination). 
Furthermore, atomic uncertainties were calculated at a frequency of 1 ps to provide insight into the reliability of the physics during the trajectory. The uncertainties observed over the course of our two simulations are provided in Figure \ref{fig:simulation}.

\begin{figure}[htbp]
    \centering
        \begin{tikzpicture}[>=latex]
    \begin{groupplot}[
            /tikz/thick,
            group style={
                group size=1 by 2,
                vertical sep=0.2cm,
                xlabels at=edge bottom,
                ylabels at=edge left,
                xticklabels at=edge bottom,
            },
            width=3.0in,
            height=1.6in,
            ylabel near ticks,
            xlabel={Simulation time [ps]},
            legend style={at={(0.0,1.0)},anchor=north west},
            legend columns=1,
            ymin=-1,
        ]
            \nextgroupplot[
                ylabel style={align=center},
                ylabel={\# reaction events},
            ]
            \addplot+  table[y index=2] {data/reactions2.dat};
            %\addlegendentry{H$_2$ recombination};
            %\addplot+  table[y index=2] {data/reactions2.dat};
            %\addlegendentry{H$_2$ splitting};
            
            \nextgroupplot[
                ylabel={Uncertainty},
                height=1.5in,
                legend style={at={(0.5,0.5)}},
                ymin=0,
                ymax=0.007,
                scaled y ticks=false,
                y tick label style={
                    /pgf/number format/fixed,
                    /pgf/number format/precision=3,
                },
            ]
            \addplot+[red]  table[x index=1, y index=2] {data/uncertainty_long.2};
            \draw[dashed,gray] (axis cs:-10,0.005) -- (axis cs:340,0.005) node[pos=1.0,anchor=south east] {Threshold};
    \end{groupplot}
\end{tikzpicture}
    \caption{Top: Number of H$_2$ recombination and dissociation reactions vs. MD simulation time. Bottom: Maximal atomic uncertainty with the MD simulation time, where the grey line is the uncertainty decision threshold for Bayesian active learning. The large-scale simulation has uncertainty well below the threshold, indicating the model is confident in its predictions.}
    \label{fig:simulation}
\end{figure}
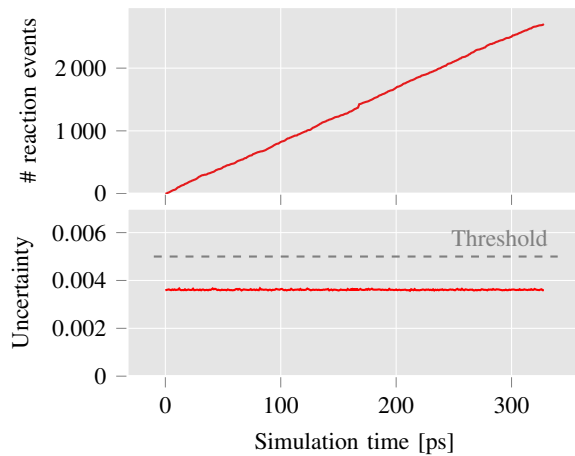

% \begin{figure}[htbp]
%     \centering
%     \includegraphics[width=0.45\textwidth]{system_unc.png}
%     \caption{Maximum atomic uncertainty as a function of simulation step across both ???B atom simulations on Summit.}
%     \label{fig:ptH_unc}
% \end{figure}

% Here, we observe ???.

% \begin{figure}[htbp]
%     \centering
%     \includegraphics[width=0.45\textwidth]{reactions.png}
%     \caption{Reactions?.}
%     \label{fig:ptH_unc}
% \end{figure}

Prior to deployment, the resulting MLFF was validated against DFT and compared with the ReaxFF model \cite{Gai16p9780}, as reported previously\cite{vandermause2021active}. The parity plots for force predictions are shown in Figure~\ref{fig:mae}. The mean absolute errors (MAE) for energies, forces, and stresses (trace) are 1.7~meV/atom, 91 \& 74~meV/\AA~(Pt \& H), and 0.6~meV/\AA$^{3}$ with our FLARE model, all of which are an order of magnitude lower than 33~meV/atom, 631 \& 676~meV/\AA~(Pt \& H), and 15.7~meV/\AA$^{3}$ obtained with the ReaxFF model, respectively.

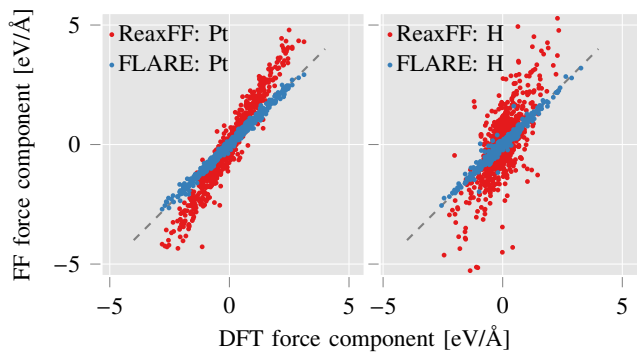
\begin{figure}[htbp]
    \centering
        \begin{tikzpicture}[>=latex]
    
    \begin{groupplot}[
            /tikz/thick,
            group style={
                group size=2 by 1,
                horizontal sep=0.2cm,
                xlabels at=edge bottom,
                ylabels at=edge left,
                xticklabels at=edge bottom,
                yticklabels at=edge left,
            },
            width=2.0in,
            height=2.0in,
            ylabel near ticks,
            xlabel={DFT force component [eV/\AA]},
            ylabel={FF force component [eV/\AA]},
            legend style={at={(0.0,1.0)},anchor=north west},
            legend columns=1,
            ymin=-5.5,
            ymax=5.5,
            axis equal,
            xlabel style={at={(1.0,-0.15)},anchor=north}
        ]

            \nextgroupplot[
            ]
            \addlegendentry{ReaxFF: Pt};
            \addplot+[only marks, mark size=0.5]  table[x index=0, y index=1] {data/ReaxFF_Pt.txt};
            \addlegendentry{FLARE: Pt};
            \addplot+[only marks, mark size=0.5]  table[x index=0, y index=1] {data/SGP_Pt.txt};
            \draw[dashed,gray] (axis cs:-4,-4) -- (axis cs:4,4); % node[pos=1.0,anchor=south east] {Threshold};
            
            \nextgroupplot[
                %ylabel={}, % not needed, can move the other one down instead
                % ylabel style={align=center},
                % ylabel={\# reaction events},
                % title={H},
                xlabel={},
            ]
            \addlegendentry{ReaxFF: H};
            \addplot+[only marks, mark size=0.5]  table[x index=0, y index=1] {data/ReaxFF_H.txt};
            \addlegendentry{FLARE: H};
            \addplot+[only marks, mark size=0.5]  table[x index=0, y index=1] {data/SGP_H.txt};
            \draw[dashed,gray] (axis cs:-4,-4) -- (axis cs:4,4);
            %\addlegendentry{H$_2$ recombination};
            %\addplot+  table[y index=2] {data/reactions2.dat};
            %\addlegendentry{H$_2$ splitting};
    \end{groupplot}
\end{tikzpicture}
    \caption{Comparison of force predictions from DFT and force fields (FLARE in blue, ReaxFF in red) predictions \cite{vandermause2021active}.}
    \label{fig:mae}
\end{figure}

% \begin{table}[htbp]
% \caption{H adsorption energies ($E_{\textrm{ads}}$) and diffusion barriers ($E_{\textrm{dif}}$) on Pt(111), compared with DFT and the FLARE model. Forward and reverse barriers are listed for each diffusion pathway \cite{vandermause2021active}.}
% \centering
% \begin{tabular}{cccc}
% \hline
% Quantity & Sites & DFT & FLARE \\ \hline
% \multirow{3}{*}{$E_{\textrm{ads}}$ (eV)} & FCC hollow & $-$0.52 & $-$0.48 \\
%  & HCP hollow & $-$0.47 & $-$0.43 \\
%  & Top & $-$0.49 & $-$0.44 \\ \hline
% \multirow{3}{*}{$E_{\textrm{dif}}$ (eV)} & FCC $\rightleftharpoons$ HCP & 0.07 \hspace{0.2cm} 0.02 & 0.05 \hspace{0.2cm} 0.01 \\
%  & FCC $\rightleftharpoons$ Top & 0.15 \hspace{0.2cm} 0.12 & 0.15 \hspace{0.2cm} 0.10 \\
%  & HCP $\rightleftharpoons$ Top & 0.09 \hspace{0.2cm} 0.11 & 0.12 \hspace{0.2cm} 0.11 \\ \hline
% \end{tabular}
% \label{tab:energetics}
% \end{table}

The full transition state pathways were examined for \ce{H2} dissociative adsorption and atomic H diffusion on Pt(111) at the low-coverage limit \cite{vandermause2021active}. The configuration energies are in excellent agreement between our FLARE model and DFT. ReaxFF is precluded from transition state modeling here, as it wrongly predicts the top site to be the most stable by 0.16~eV compared to the HCP hollow site.

The ultimate accuracy performance test for reactive MD simulations is the comparison with experimental measurements of reaction rates. In that regard, our first-iteration H/Pt FLARE model produced an estimate of the effective exchange activation energy of 0.25 eV in excellent agreement with the experimentally measured value of 0.23 eV \cite{vandermause2021active}. This indicates that the FLARE MD simulation accurately captures all the steps of the catalytic reaction chain and retains first-principles accuracy.

\subsection{Systems and Environment for Measurement of Performance}

All simulations were performed using the Kokkos portability library in conjunction with LAMMPS MD.

All of our scaling and production MD simulations were performed on Summit, the GPU-based, leadership-class HPC system at Oak Ridge National Laboratory. Summit has 4608 nodes, each with 6 NVIDIA V100 GPUs and an IBM Power9 CPU.

We used the LAMMPS version \texttt{stable\_29Sep2021} \texttt{\_update2} with our FLARE extension. The code was compiled with GCC 9.3, CUDA 11.0 and Spectrum MPI 10.4, using CUDA-aware MPI for efficient GPU-GPU communication.

\subsection{Measurement metrics}
We consider the overall time-to-solution from first principles defined in section \ref{sec:sota} as the key measurement metric: it is the sum of training data generation, force fields training, and production time.
For the first two steps, we simply measure the wall time required to run the different Bayesian active learning simulations, and then combine the resulting training data into one complete model. We compare the wall time of the active learning run with the time estimate of using AIMD for every step.

% \begin{figure}[h]
%     \centering
%     \includegraphics[width=0.48\textwidth]{MAE.pdf}
%     \caption{Parity plot of the predicted force components against DFT values for Pt and H. The parity lines are shown in black. Predictions of the SGP model and the ReaxFF model are shown in blue and red, respectively. The mean absolute error (MAE) values are indicated in the legend.}
%     \label{fig:MAE}
% \end{figure}

For the final molecular dynamics, we perform extensive scaling experiments for a wide range of system sizes and computational resources, to demonstrate close to the ideal linear scaling of MD. We measure strong scaling by increasing the number of nodes for a fixed system size and weak scaling by maintaining a constant per-node system size. Each simulation is run for 2 minutes (excluding the initial setup) to ensure sufficient sampling. We report the results in the unit of mega-atom-steps per second per node, as this is a convenient measure that should be constant for all simulations with linear scaling. Finally, we track the fraction of the simulation time actually spent inside FLARE, to see the impact of communication, neighbor lists etc. varies. This information is provided by the internal timing mechanisms of LAMMPS.

%system and environment where performance was measured (1 p max)

\section{Performance Results}
%\textit{include scalability (weak and strong), time to solution, efficiency (of bottleneck resources), and peak performance}
In this section, we provide details of the SOTA performance of FLARE on scalability/production and training from first principles. In Table \ref{tab:radar} we compared FLARE with the two other MLFF models used for large-scale MD simulations on Summit.

%\begin{figure}[htbp]
%    \centering
%    \includegraphics[width=0.45\textwidth]{radar.pdf}
%    \caption{Comparative analysis of state of the art force fields. PtH model accuracy is quantitatively available for FLARE and ReaxFF \cite{vandermause2021active}, but not for SNAP and DeePMD.}
%    \label{fig:radar}
%\end{figure}

\begin{table}[h]
    \caption{Comparative analysis of SOTA MLFF models on Summit. Production speed unit:  M\,atom-steps/s/node}
    \label{tab:radar}
    \centering
\begin{tabular}{cccccc}
\hline
  \multirow{2}{*}{Model} & \multirow{2}{*}{Atoms} & \multirow{2}{*}{Speed}  & \multirow{2}{*}{\makecell{Production scale \\ uncertainty}} & \multirow{2}{*}{\makecell{Active \\ learning}} \\
  %& atoms & nodes  \\
  \\
 \hline
DeePMD & 3.9 B  & 2.0 & \cellcolor{salmon}No & \cellcolor{grannysmith}Yes \\
SNAP & 20 B  & 6.21 & \cellcolor{salmon}No & \cellcolor{salmon}No \\
%ReaxFF & 1 M & 1/3 & 1.87 & \cellcolor{salmon}No & \cellcolor{salmon}No \\
FLARE & \textcolor{flarered}{\textbf{500 B}}  & \textcolor{flarered}{\textbf{10.5}} & \cellcolor{grannysmith}Yes & \cellcolor{grannysmith}Yes \\
  \hline
    \end{tabular}
\end{table}
\subsection{Scalability\label{subsec:scalability}}
We performed strong scaling experiments with system sizes from \(10^6\) to \(10^{11}\) atoms, with up to 4096 nodes (24576 GPUs) on Summit in powers of 2. We also performed one benchmark with 0.5 \emph{trillion} atoms, which is by far the largest MLFF benchmark ever performed and required 4556 nodes. The \(10^9\), \(10^{10}\) and \(10^{11}\) atom simulations needed 8, 128 and 1024 nodes to fit in GPU memory, respectively, when FLARE was limited to use up to 2 GB for its batching of atoms.  When going beyond \(10^{10}\) atoms, we found that the initial, CPU-based replication of the 245200-atom structure became too expensive. Instead, the \(10^{11}\) benchmarks were initiated using a faster protocol of creating a lattice of 272-atom unit cells, which resulted in a structure with a slightly different Pt surface H coverage value.

The strong scaling results are shown in Figure \ref{fig:strongscaling}. In the top panel, the speed of FLARE is compared to the peak performance of SNAP~\cite{nguyen2021billion} and DeePMD~\cite{guo2022extending}. FLARE outperforms SNAP and DeePMD by factors of 1.7 and 5.3, respectively, when in the regime of linear scaling, achieving a peak performance of 10.5 M\,atom-steps/s/node. For our chosen PtH system, we observe that FLARE leaves the linear scaling regime earlier than for a simple bulk system like silicon. In the bottom panel, we measure the fraction of time spent inside FLARE during the strong scaling simulations. The deviation from 100~\% aligns perfectly with the deviation from linear scaling and demonstrates the impact of communication and load imbalance. Communication and load imbalance are much more severe issues when running complicated heterogeneous, near-two-dimensional systems like PtH than when running simple bulk crystals.

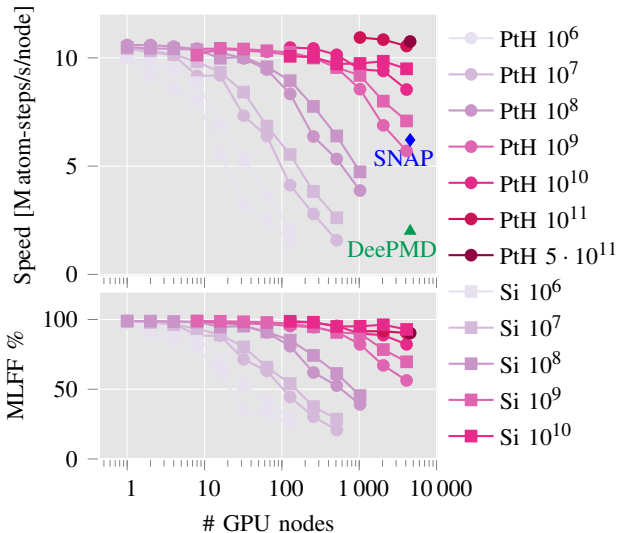
\begin{figure}[htbp]
    \centering
        \begin{tikzpicture}[>=latex]
        \pgfplotsset{cycle list/PuRd-8}
    \begin{groupplot}[
            /tikz/thick,
            group style={
                group size=1 by 2,
                vertical sep=0.2cm,
                xlabels at=edge bottom,
                ylabels at=edge left,
                xticklabels at=edge bottom,
            },
            width=2.4in,
            height=2.0in,
            ylabel near ticks,
            xlabel={\# GPU nodes},
            legend style={at={(1.0,1.0)},anchor=north west},
            legend columns=1,
            xmode=log,
            log basis x=10,
            log10 x ticks with fixed point,
            ymin=-0.1,
            clip=false,
        ]
            \nextgroupplot[
                ylabel={Speed [M\,atom-steps/s/node]},
            ]
            \pgfplotsset{cycle list shift=1}
            \addplot+[mark=*]  table {data/pthstrong2.dat};
            \addlegendentry{PtH \(10^6\)};
            \addplot+[mark=*]  table {data/pthstrong6.dat};
            \addlegendentry{PtH \(10^7\)};
            \addplot+[mark=*]  table {data/pthstrong20.dat};
            \addlegendentry{PtH \(10^8\)};
            \addplot+[mark=*]  table {data/pthstrong64.dat};
            \addlegendentry{PtH \(10^9\)};
            \addplot+[mark=*]  table {data/pthstrong202.dat};
            \addlegendentry{PtH \(10^{10}\)};
            \addplot+[mark=*]  table {data/pthstrong19175.dat};
            \addlegendentry{PtH \(10^{11}\)};
            \addplot+[mark=*] coordinates {(4556, 10.755)};
            \addlegendentry{PtH \(5\cdot10^{11}\)};
            \pgfplotsset{cycle list shift=-6}
            \addplot+[mark=square*]  table {data/sistrong50.dat};
            \addlegendentry{Si \(10^6\)};
            \addplot+[mark=square*]  table {data/sistrong108.dat};
            \addlegendentry{Si \(10^7\)};
            \addplot+[mark=square*]  table {data/sistrong232.dat};
            \addlegendentry{Si \(10^8\)};
            \addplot+[mark=square*]  table {data/sistrong500.dat};
            \addlegendentry{Si \(10^9\)};
            \addplot+[mark=square*]  table {data/sistrong1077.dat};
            \addlegendentry{Si \(10^{10}\)};
            %\addplot+[mark=square*]  table {data/snap1e6.dat};
            %\addlegendentry{SNAP \(10^6\)};
            %\addplot+[mark=square*]  table {data/snap1e7.dat};
            %\addlegendentry{SNAP \(10^7\)};
            %\addplot+[mark=square*]  table {data/snap1e8.dat};
            %\addlegendentry{SNAP \(10^8\)};
            %\addplot+[mark=square*]  table {data/snap1e9.dat};
            %\addlegendentry{SNAP \(10^9\)};
            %\addplot+[mark=square*]  table {data/snap2e10.dat};
            %\addlegendentry{SNAP \(2\cdot10^{10}\)};
            %\pgfplotsset{cycle list shift=-7}
            \addplot+[blue,only marks,mark=diamond*] coordinates {(4560, 6.21)};
            \draw[blue] (axis cs:4560,6.21) node[anchor=70] {SNAP};
            \addplot+[ForestGreen,only marks,mark=triangle*] coordinates {(4560, 2.0)};
            \draw[ForestGreen] (axis cs:4560,2.0) node[anchor=45] {DeePMD};
            %\addlegendentry{DeepMD \(3.4\cdot10^{9}\)};
            %\draw[dashed,gray] (axis cs:1,6.21) -- (axis cs:5500,6.21) node[pos=0.0,anchor=north west] {SNAP};
            %\draw[dashed,gray] (axis cs:1,2.0) -- (axis cs:5500,2.0) node[pos=0.0,anchor=north west] {DeePMD};
            
            \nextgroupplot[
                ylabel={MLFF \%},
                height=1.5in,
                legend style={at={(1.0,0.3)},anchor=west},
                ymax=120,
            ]
            \pgfplotsset{cycle list shift=1}
            \addplot+[mark=*]  table[y index=2] {data/pthstrong2.dat};
            \addplot+[mark=*]  table[y index=2] {data/pthstrong6.dat};
            \addplot+[mark=*]  table[y index=2] {data/pthstrong20.dat};
            \addplot+[mark=*]  table[y index=2] {data/pthstrong64.dat};
            \addplot+[mark=*]  table[y index=2] {data/pthstrong202.dat};
            \addplot+[mark=*]  table[y index=2] {data/pthstrong19175.dat};
            \addplot+[mark=*] coordinates {(4556, 90.27)};
            \pgfplotsset{cycle list shift=-6}
            \addplot+[mark=square*]  table[y index=2] {data/sistrong50.dat};
            \addplot+[mark=square*]  table[y index=2] {data/sistrong108.dat};
            \addplot+[mark=square*]  table[y index=2] {data/sistrong232.dat};
            \addplot+[mark=square*]  table[y index=2] {data/sistrong500.dat};
            \addplot+[mark=square*]  table[y index=2] {data/sistrong1077.dat};
    \end{groupplot}
\end{tikzpicture}
    \caption{Strong scaling from \(10^6\) to \(5\cdot10^{11}\) atoms. Top: Speed comparison with the peak performance for SNAP (\(2\cdot10^{10}\) atoms) and DeePMD (\(3.4\cdot10^{9}\) atoms). Bottom: The percentage of the simulation spent computing forces.}
    \label{fig:strongscaling}
\end{figure}

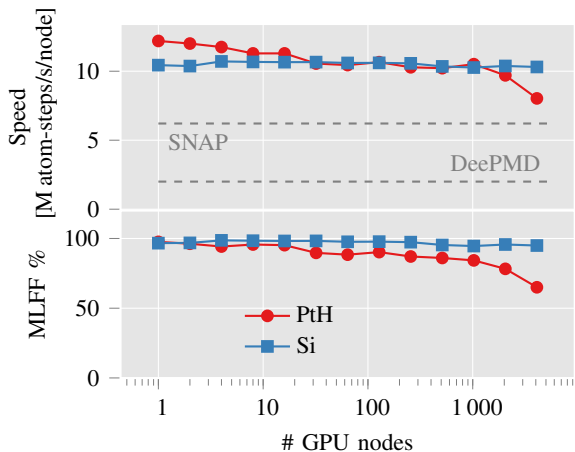
\begin{figure}[htbp]
    \centering
        \begin{tikzpicture}[>=latex]
    \begin{groupplot}[
            /tikz/thick,
            group style={
                group size=1 by 2,
                vertical sep=0.0cm,
                xlabels at=edge bottom,
                ylabels at=edge left,
                xticklabels at=edge bottom,
            },
            width=3.0in,
            height=1.6in,
            ylabel near ticks,
            xlabel={\# GPU nodes},
            legend columns=1,
            xmode=log,
            log basis x=10,
            log10 x ticks with fixed point,
            ymin=-0.1,
            clip=false,
        ]
            \nextgroupplot[
                ylabel style={align=center},
                ylabel={Speed\\{} [M\,atom-steps/s/node]},
            ]
            %\addlegendentry{Si};
            \addplot+[mark=*]  table {data/pthweak.dat};
            \addplot+[mark=square*]  table {data/siweak.dat};
            %\addlegendentry{PtH};
            \draw[dashed,gray] (axis cs:1,6.21) -- (axis cs:5500,6.21) node[pos=0.0,anchor=north west] {SNAP};
            \draw[dashed,gray] (axis cs:1,2.0) -- (axis cs:5500,2.0) node[pos=1.0,anchor=south east] {DeePMD};
            
            \nextgroupplot[
                ylabel={MLFF \%},
                height=1.5in,
                legend style={at={(0.5,0.5)}},
                ymax=120,
            ]
            \addplot+[mark=*]  table[y index=2] {data/pthweak.dat};
            \addlegendentry{PtH};
            \addplot+[mark=square*]  table[y index=2] {data/siweak.dat};
            \addlegendentry{Si};
    \end{groupplot}
\end{tikzpicture}
    \caption{Weak scaling from 1 to 4096 nodes. Top: Speed comparison with SNAP and DeePMD. Bottom: The percentage of the simulation spent computing forces.}
    \label{fig:weakscaling}
\end{figure}

Figure \ref{fig:weakscaling} shows the weak scaling of FLARE by maintaining approximately \(10^7\) atoms per node and scaling from 1 to 4096 nodes. We compare the weak scaling of a simple crystal system, silicon, to that of our heterogeneous, catalytic system. We find near-perfect weak scaling in both cases, but the slightly improved scaling of the silicon simulations demonstrates the effect of load imbalance for the catalytic system and its heterogeneous geometry. Based on these observations, we decided to continuously perform time-weighted re-balancing of the spatial decomposition for long-timescale production runs.

\subsection{Real-application performance with uncertainties\label{big-md}}
To establish real-application performance, we ran a short production simulation with 10 billion atoms on 4556 nodes, which included calculation and storage of atomic positions, hydrogen coordination numbers, and predicted variance of all atoms.

We achieve a sustained performance of 8.7 M\,atom-steps/s/node, which is 83 \% of the peak performance observed in the strong and weak scaling tests that do not include I/O, uncertainties or other analysis.
In particular, we compute the uncertainties of all the atoms on-the-fly, making this the first micrometer scale Bayesian molecular dynamics simulation. By monitoring the maximum uncertainty, we are able to verify that the simulation does not encounter unfamiliar configurations or produce any nonphysical behavior, remaining in the configuration space for which the MLFF is confident and reliable.

To confirm that our large-scale models are producing the expected surface reactions, we show on Figure \ref{fig:simulation} the number of reaction events (hydrogen molecule dissociations and recombinations) as a function of time for an example simulation with \(10^6\) atoms on 16 nodes. The final simulation time is 320 ps, providing a much larger number of reaction events and statistics compared to our previous work~\cite{vandermause2021active}. We also verify that the uncertainties remain low throughout the simulation.

\begin{figure}[htbp]
    \centering
    \includegraphics[width=0.48\textwidth]{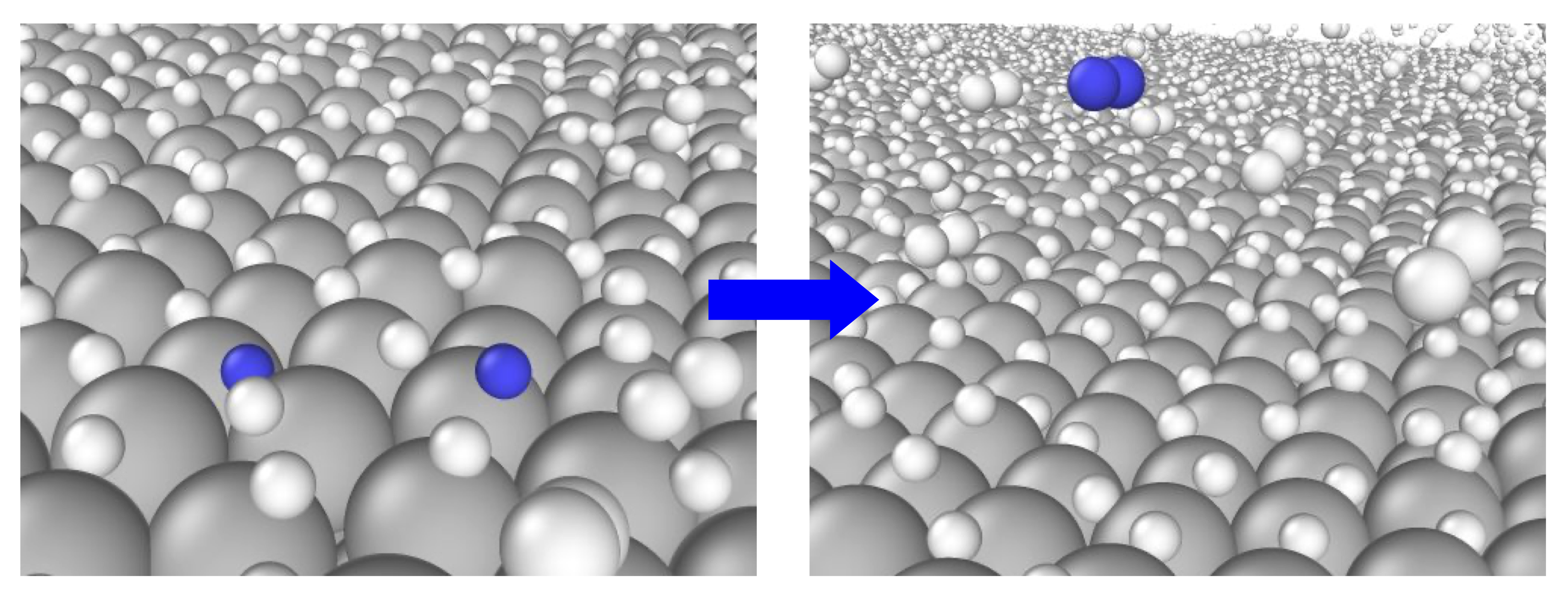}
    \caption{Snapshots of the reaction event of H$_2$ recombination. Left: before the recombination. Right: after the recombination.}
    \label{fig:big-simulation}
\end{figure}

\subsection{Time to solution from first principles}

As introduced in Section\ref{sec:innovation}, FLARE utilizes the quantitative principled uncertainty of the Gaussian process model to realize the Bayesian active learning workflow, which facilitates exploration of configuration space with a fast surrogate model while monitoring rare events by uncertainty. Meanwhile, due to the high data efficiency of the kernel-based regression \cite{vandermause2020fly}, only a small number of DFT calculations are needed to converge the model to a satisfactory accuracy.
For example, as shown in Table \ref{tab:training_stats}, Bayesian active learning only makes 575 DFT calls for different configurations at different conditions, while the same exploration trajectory would take 138.7 years to generate with AIMD. In comparison, training a model like SNAP/HDNNP with non-accelerated AIMD data usually requires more computational resources and explores orders of magnitude shorter time-scales.
FLARE also demonstrates a $13\times$ higher data efficiency for learning the H/Pt heterogeneous system model, compared to the DP-GEN active learning scheme of DeePMD for pure copper \cite{zhang2020dp}. 

\begin{table}[htbp]
    \caption{Training set collected by Bayesian active learning; MD temperature $T$; total simulation time $\tau_{\text{sim}}$; total simulation and training wall time $\tau_{\text{wall}}$ (hours); number of CPU nodes $\tau_{\text{nodes}}$ used; number of atoms in the simulation $N_{\text{atoms}}$; total number of DFT calls made during training $N_\text{DFT}$; node time needed to generate an equivalent ab-initio molecular dynamics trajectory without active learning $\tau_\text{AIMD}$.  ``Master'' refers to the offline training of a master potential for production that combines all the training data from Bayesian active learning. $^{*}$ Active-learning workflow started with a pre-trained model.}
    \label{tab:training_stats}
    \centering
\begin{tabularx}{0.48\textwidth}{ 
   >{\raggedright}X 
   >{\centering}X
   >{\centering}X
   >{\centering}X
   >{\centering}X
   >{\centering}X
   >{\centering}X
   >{\centering\arraybackslash}X }
 \hline
 \hline
 System & $T$ (Kelvin) & $\tau_{\text{sim}}$ (ps) & $\tau_{\text{wall}}$ (hours) & $N_\text{nodes}$ &  $N_{\text{atoms}}$ & $N_\text{DFT}$ & $\tau_\text{AIMD}$ (years) \\ 
 \hline
 H$_2$ & 1500 & 5.0 & 1.2 & 1 & 54 & 24 & 4.1 \\ 
 H$_2^{*}$ & 2100 & 10.0 & 3.5 & 2 & 108 & 2 & 8.6 \\
 Pt(111) & 300 & 4.0 & 1.2 & 2 & 54 & 4 & 3.5 \\ 
 Pt & 1500 & 10.0 & 1.7 & 2 & 108 & 6 & 3.4 \\ 
 H/Pt & 1500 & 3.7 & 61.4 & 2 & 73 & 216 & 75.0 \\
 H/Pt$^{*}$ & 2100 & 26.5 & 143.1 & 3 & 92 & 323 & 44.1 \\
 Master & - & - & 2.7 & 1 & - & - & - \\
 \hline
 Total & - & 59.2 & 214.8 & - & - & 575 & 138.7 \\
 \hline
\end{tabularx}
\end{table}

Quantitative uncertainty of each local environment inherent in the the Bayesian force field allows the models to be improved in accuracy and robustness in hierarchical fashion, even using uncertainty of large-scale production runs. 
As an example, the initial force field \cite{vandermause2021active}, trained on bulk Pt at T=1500K, Pt(111) at 300K, \ce{H2} at 1500K, and \ce{H2}/Pt(111) at 1500K, was deployed in 0.5M atom MD simulations for 24 hours on Longhorn during the Texascale Day at TACC. 
In the large-scale simulation, suspicious non-physical phenomena were captured by a spike in the uncertainty. Visual inspection revealed Pt desorption from the surface and clustering of H atoms, meaning that the attractive and repulsive regions of the Pt and H potentials were not sufficiently sampled during training with the small system size.
Subsequently, two additional training runs with Bayesian active learning were employed, targeting local configurations observed in large-scale MD runs that had high uncertainty.
For improving the repulsive interaction of H-H, we augmented the training set with high-pressure \ce{H2} gas with twice the density and a higher temperature of 2100 K. For the attractive interaction of Pt-Pt, we also included data on H/Pt system at 2100 K, allowing the Pt slab to melt over the course of $\sim$27 ps. 

% We can compare the Bayesian active learning workflow directly to AIMD, which requires DFT to be performed at every time-step. As shown in Table \ref{tab:training_stats}, the Bayesian active learning is set up on the Pt(111), Pt bulk, H2 gas, and H/Pt combined system surveying a total of tens of pico-seconds of dynamics.
% The entire process of Bayesian active learning, including the exploration of the phase space, the DFT calculation and the training of the model only takes days starting from scratch, with each configuration running distributedly on one or two CPU nodes, while the same trajectory can take a century to generate in DFT.

\section{Implications}
%(\textit{implications for future systems and applications})

%\subsection{Increase in accessible length-scales for materials property prediction}
Simulating dynamics of complex reactive systems on atomistic level is a grand challenge in computational materials science and chemistry. The FLARE method, when deployed on leadership-class resources, opens possibilities to simulate thermal and reactive dynamics at previously inaccessible length and time scales while maintaining first-principles accuracy. We have demonstrated excellent scaling to hundreds of billions of atoms for a heterogeneous reactive system, decreased total time-to-solution from first principles, and uncertainty quantification enabling more robust and principled MLFF generation and deployment. Importantly, FLARE offers a dramatic step forward in the direct comparison of accurate reactive MD simulations to experimental observations, enabled by the efficient scaling achieved by these MLFFs to billion-atom system sizes. By employing fast prediction and uncertainty quantification, the workflow can be applied to learn models and simulate systems with increasing chemical and structural complexity, including alloys, biological enzymes, polymers, electrode interfaces and, notably, heterogeneous catalysis. In metallic alloys simulations with billions of atoms can help answer long-standing questions about fatigue and grain boundary segregation that govern mechanical properties of structural materials. Computations of ionic and thermal currents can be used to estimate non-equilibrium transport properties in liquid and solid-state batteries and realistic thermoelectric materials, in the presence of interfaces and nontrivial microstructure. Understanding heterogeneous phenomena such as shock waves and crack propagation are also in particular need of large-scale fast MD simulations.

 Reactive processes can benefit greatly from FLARE at scale, especially those that have complicated reaction networks, where multiple mechanisms may exist but are not known. Fast MD simulations may be used to reveal these mechanisms, while at the same time learning the relevant energies and forces. In the case demonstrated in this work, catalytic reaction mechanisms of H$_2$ interacting with the Pt surface at high coverage are complicated and were discovered automatically by the active learning algorithm, which led to the highly accurate prediction of overall reaction rates. Robust determination of the mechanisms could lead to more efficient material design, e.g. new catalysts for methane activation to biofuels and other industrial reactions. In these processes, the structure of the active site is critically important, but may be exhibit dynamic restructuring in response to changing reaction conditions. Catalytic processes like these present excellent opportunities for FLARE as they involve complicated reaction mechanisms over a range of timescales, which can be discovered using active learning, bypassing the need for any \textit{a priori} assumptions or expert chemical intuition. This advantage signals a departure from previous efforts relying on chemical expertise to manually enumerate elementary reaction mechanisms and perform kinetic analyses with prior assumptions of rate-limiting steps. We emphasize that progress in deciphering and simulating reactions necessarily requires both fast and scalable MD and uncertainty quantification needed to efficiently explore the configuration space and obtain reliable large-scale models.

In addition to the scientific impacts, a notable practical advantage of uncertainty-driven active learning is that it represents a step towards full automation, allowing a broader community of scientists to approach chemical or material property predictions from first principles. Users without expert chemical intuition would be able to train a MLFF automatically, and perform uncertainty-aware molecular dynamics. This process is drastically accelerated with the inclusion of active learning in model creation, decreasing the total time-to-solution from inception of the scientific goal to material property prediction from MD at scale. Where previously it took months to develop force fields that were not guaranteed to be accurate, with FLARE the models can take a few days to train without human input and provide internal reliability metrics.

Efficient parallel implementation enabling full GPU utilization for both small and large systems opens the possibility to use MD to study extremely long time scales in smaller systems, as well as to explore extremely large systems, with up trillion-atom scale on leadership-class machines. Due to the adoption of the Kokkos GPU performance portability library, we ensure long-lasting impact, as the optimizations of our models will translate across future GPU architectures adopted for supercomputers, including NVIDIA, Intel and AMD hardware, without requiring low-level code redevelopment.

FLARE is made available as open-source software under the MIT license, which encourages adoption by academic and commercial research and development efforts. Multiple research groups in universities, national labs and industry (representative letters attached) are already adopting the new MLFF models to perform large-scale accurate MD for a variety of real-world applications, including understanding and design of advanced semiconductor and catalytic materials.

Finally, the applications of the fast scalable FLARE models are not limited to molecular dynamics. Combining them with grand-canonical Monte Carlo simulations will allow for rapid explorations of phase diagrams of complex alloys. Metadynamics, umbrella sampling and other enhanced sampling techniques based on the new MLFF can be used to estimate free energy surfaces and reaction rates. In the domain of soft and biological materials, coarse grained models can readily be developed using the FLARE active learning and force matching formalism, allowing for even longer time scales to be examined.

\section{Acknowledgement}
We would like to acknowledge helpful discussions with Stan Moore from Sandia Natl. Lab. Computational resources on Summit were provided by the Oak Ridge Leadership Computing Facility. The authors acknowledge computing resources provided by the Harvard University FAS Division of Science Research Computing Group and by the Texas Advanced Computing Center (TACC) at The University of Texas at Austin.
% under allocation DMR20013??. 
%J.S.L. and L.S. were supported by Integrated Mesoscale Architectures for Sustainable Catalysis (IMASC), an Energy Frontier Research Center funded by the US Department of Energy, Office of Science, Office of Basic Energy Sciences under Award No. DE-SC0012573. C.J.O. is supported by the National Science Foundation Graduate Research Fellowship Program, Grant No. DGE1745303.
%We acknowledge support from the US Department of Energy (DOE), Office of Science, Office of Basic Energy Sciences (BES) under Award No. DE-SC0020128 and Award No. DE-SC0022199, and the National Science Foundation under Grant No. 2003725.

\def\url#1{}
\def\doi#1{}
\bibliographystyle{IEEEtran}
\bibliography{main}

\end{document}